\documentclass[12pt]{article}

\usepackage{amsthm,amsfonts,amsmath,braket,algorithm,graphicx,authblk}

\usepackage{fullpage}

\theoremstyle{definition} 
\theoremstyle{definition} 
\newtheorem {theorem} {Theorem}

\newcommand{\kb}[1]{\ket{#1}\bra{#1}}
\newcommand{\bk}[1]{\braket{#1|#1}}

\newcommand{\msg}[1]{\texttt{`#1'}}

\newcommand{\MR}{\texttt{Measure-Resend}}
\newcommand{\R}{\texttt{Reflect}}
\newcommand{\modeflip}{\texttt{Mode $=$ FLIP}}
\newcommand{\modenoflip}{\texttt{Mode $=$ NO-FLIP}}
\newcommand{\mode}{\texttt{Mode}}

\newcommand{\st}{\text{ } : \text{ }}

\floatname{algorithm}{Protocol}

\title{Mediated Semi-Quantum Key Distribution with Improved Efficiency}

\author[1]{Julia Guskind}
\author[1]{Walter O. Krawec\footnote{Email: \texttt{walter.krawec@uconn.edu}}}
\affil[1]{\small{Department of Computer Science and Engineering}\\\small{University of Connecticut}\\\small{Storrs, CT 06269 USA}}

\date{}

\begin{document}

\maketitle

\begin{abstract}
Mediated semi-quantum key distribution involves the use of two end-users who have very restricted, almost classical, capabilities, who wish to establish a shared secret key using the help of a fully-quantum server who may be adversarial.  In this paper, we introduce a new mediated semi-quantum key distribution protocol, extending prior work, which has asymptotically perfect efficiency.  Though this comes at the cost of decreased noise tolerance, our protocol is backwards compatible with prior work, so users may easily switch to the old (normally less efficient) protocol if the noise level is high enough to justify it.  To prove security, we show an interesting reduction from the mediated semi-quantum scenario to a fully-quantum entanglement based protocol which may be useful when proving the security of other multi-user QKD protocols.
\end{abstract}

\section{Introduction}

Quantum key distribution (QKD) allows two parties, who we refer to customarily as Alice and Bob, to establish a shared secret key.  Unlike classical communication protocols, where security must always depend on unproven computational assumptions, QKD is secure against computationally unbounded adversaries (i.e., adversaries who are bounded only by the laws of quantum physics).  One fascinating question in this field of research is, ``how quantum'' must a protocol be to gain this advantage over classical protocols?  To this end, the notion of \emph{semi-quantum cryptography} was introduced in 2007 by Boyer et al., \cite{SQKD-first}.  Originally only for key distribution \cite{SQKD-second,zou2009semiquantum,zhang2017fault,tsai2018semi,amer2019semiquantum,tsai2019cryptanalysis,iqbal2020high}, the field of semi-quantum cryptography has advanced to other applications including secure direct communication \cite{SDC-1,SDC-2,SDC-3,SDC-4}, secret sharing \cite{secret-1,secret-2,secret-3}, identity verification \cite{ident-1,ident-2}, and private state comparison \cite{priv-comp-1,priv-comp-2}.  See \cite{SQKD-survey} for a recent survey on semi-quantum research and \cite{QKD-survey1,QKD-survey2} for general surveys on quantum cryptography.

Semi-quantum protocols involve at least one party who is semi-quantum or ``classical'' in nature.  Such a party is only able to interact with the quantum channel in a limited, almost classical, way.  In particular, this party can only measure and send in a single publicly known basis (generally the computational $Z$ basis spanned by $\ket{0}$ and $\ket{1}$).  Typically, the party may measure in this basis, observing outcome $\ket{r}$, and then resend the result $\ket{r}$, an operation known as $\MR$, though some recent implementations designed for use with practical devices require only $Z$ basis measurements and not state preparation \cite{boyer2017experimentally,krawec2018practical,massa2019experimental,silva2021semi} (with \cite{silva2021semi} being also secure in a strong semi-device independent security model).  Beyond this, the party may also choose to ``disconnect'' from the quantum channel, letting any signal pass through their lab undisturbed back to the original sender (an operation denoted $\R$, as it ``reflects'' the signal back to the sender).  Permutation of a signal is also allowed \cite{SQKD-second}, though we do not require this operation here.  Notice that if all parties were restricted in this manner, the protocol would be mathematically equivalent to a purely classical communication protocol and, thus, unconditional security would be impossible.  Semi-quantum cryptography, therefore, seeks to better understand this ``gap'' between classical and quantum secure communication.

In 2015, it was shown that key-distribution is possible when \emph{both} parties, Alice and Bob, are semi-quantum according to the above definition, so long as a fully-quantum third party server is available \cite{krawec2015mediated}.  Interestingly, security is possible even when this server is actually the adversary.  Such a protocol is called a \emph{mediated semi-quantum key distribution} (M-SQKD) protocol.

Since this original M-SQKD protocol, several other mediated and multi-user semi-quantum protocols have been developed with various advantages and disadvantages.  In general three avenues of research exist, often with many protocols advancing more than one of these simultaneously.  First, is decreasing the necessary resources placed on the end-users; second is decreasing the necessary resources for the quantum server; and third is to increase efficiency or noise tolerance.

Some attempts have been made to decrease the resources placed on the end-users.  For instance, \cite{m-2-liu2018mediated} designed a M-SQKD protocol which did not require users to measure (though recently in \cite{m-6-zou2020three} some attacks have been found against this protocol).  An M-SQKD protocol where users did not have to prepare quantum states was proposed in \cite{massa2019experimental} along with a finite-key security analysis and an experimental proof-of-concept.  Towards reducing the server requirements, \cite{m-3-lin2019mediated} required the server to only send single qubit states, one to each user (as opposed to creating an entangled Bell state as in the original 2015 protocol).  This was followed by a Bell measurement.  Though, in \cite{m-11-lu2020collective} some attacks were shown against this protocol but with a proposed improvement.  Another protocol in \cite{m-9-chen2021efficient} requires only single qubit preparation and measurement for the server, though also requires a cycle topology (allowing the qubit to travel from the server to Alice, to Bob, then back to the server).  Finally, towards increasing noise tolerance and/or efficiency, generally new protocols are developed, or new models such as the multi-mediated SQKD model which use multiple servers to gain advantages in noise tolerance \cite{krawec2019multi} (though usually at the cost of efficiency).

Our work attempts to improve efficiency by extending the original M-SQKD protocol of \cite{krawec2015mediated} in a way that does not require additional quantum capabilities for either the server or the users.  In fact, our protocol is backwards compatible with the original 2015 M-SQKD protocol. Our extension, though potentially doubling  efficiency, comes at the cost to noise tolerance as we demonstrate later. Since our extension does not require additional quantum complexity, end-users may decide (even after the protocol is run), to execute the original M-SQKD protocol or our proposed extension here (since only the classical parts of the protocol are changed).  Taken as a whole, therefore, our work can only increase the efficiency of the 2015 M-SQKD protocol (at low noise levels), or maintain the original efficiency (at high noise levels) without sacrificing noise tolerance.  For ``low'' noise levels, one may use our extension; once the noise level is passed a certain threshold (which can be found through our key-rate bound as we show later), our extension may be deactivated, switching to the original protocol and its higher noise tolerance.  Furthermore, this is the first M-SQKD protocol \emph{with provable security} that allows, asymptotically, all communication rounds to contribute to the raw key.  While other M-SQKD protocols have also been proposed with asymptotically perfect efficiency \cite{m-2-liu2018mediated,m-3-lin2019mediated,m-9-chen2021efficient,m-10-hwang2020mediated}, they are only proven secure against certain classes of attacks.  Towards proving our protocol secure, we also show a novel reduction from this M-SQKD scenario to an entanglement based protocol.  This reduction method may be useful in analyzing other multi-user (S)QKD protocols.

We make several contributions in this paper.  First, we revisit the original M-SQKD protocol of \cite{krawec2015mediated} in order to improve its efficiency.  In particular, we extend that protocol so that, asymptotically, all rounds lead to contributions towards the distilled key (in the original protocol of \cite{krawec2015mediated}, only half the rounds could contribute asymptotically in ideal conditions).  To prove security, we rely on alternative classical post processing methods, especially mismatched measurement analysis \cite{QKD-Tom-First,QKD-Tom-KeyRateIncrease,QKD-Tom-KeyRateMismatchedDistill} (shown to be vital for many semi-quantum protocols \cite{SQKD-survey}), to bound Eve's information in this case.

Perhaps our main contribution, however, is that we devise a general proof of security for a mediated semi-quantum protocol by developing a novel reduction to an entanglement based protocol.  Due to the two-way quantum channel, standard mathematical tools used in QKD security proofs often cannot be directly applied to semi-quantum scenarios.  In this work, we show for the first time that mediated SQKD protocols may be reduced to equivalent entanglement based versions thus opening up the possibility of more rigorous analytical methods.  So far, reductions are only known for a certain subset of two-party SQKD protocols \cite{krawec2018key,iqbal2020high} but none were known for M-SQKD protocols.  Our new reduction may be highly useful to other researchers of multi-user (S)QKD protocols, providing new methods to reduce their analysis to one-way entanglement based protocols for which many mathematical tools exist to help prove them secure.

\subsection{Preliminaries}

Given a bipartite quantum state $\rho_{AB}$, we write $H(AB)_\rho$ to mean the von Neumann entropy of $\rho_{AB}$.  We write $H(A|B)_\rho$ to mean the conditional von Neumann entropy, namely $H(A|B)_\rho = H(AB)_\rho - H(B)_\rho$, where $H(B)_\rho$ is the entropy in the resulting system $\rho_B = tr_A\rho_{AB}$ after tracing out $A$.  If $\rho_{AB}$ is a classical system, then $H(A|B)_\rho$ is actually the Shannon entropy and in this case, we will often simply write $H(A|B)$  so there is no ambiguity in the fact we were discussing a classical system at that point.  We denote by $h(x)$ to be the binary Shannon entropy, namely $h(x) = -x\log_2 x - (1-x)\log_2 (1-x)$.  Finally, we denote by $\ket{\phi_i}$ to be the four Bell states, namely $\ket{\phi_0} = \frac{1}{\sqrt{2}}(\ket{00} + \ket{11})$, $\ket{\phi_1} = \frac{1}{\sqrt{2}}(\ket{00} - \ket{11})$, $\ket{\phi_2} = \frac{1}{\sqrt{2}}(\ket{01} + \ket{10})$, and $\ket{\phi_3} = \frac{1}{\sqrt{2}}(\ket{01} - \ket{10})$.

All quantum key distribution protocols, semi-quantum or otherwise, consist of two general steps.  First is the quantum communication stage which utilizes the quantum channel and the authenticated classical channel to establish a \emph{raw key}.  This is a classical bit string held by Alice and Bob which is partially correlated and partially secret.  Due to these reasons, the raw key cannot be used immediately for other cryptographic purposes.  Instead, a second classical postprocessing stage must be run which consists of an error correction protocol followed by privacy amplification.  See \cite{QKD-survey1,QKD-survey2} for more information on these standard processes.

Privacy amplification involves running the error corrected raw key through a two-universal hash function.  If the raw key size is $N$ bits long, the secret key will be of size $\ell(N) \le N$ bits.  The more information Eve has on the raw key, the smaller $\ell(N)$ will be.  A statistic of importance in QKD security analyses is its \emph{key rate}, namely the ratio $\ell(N)/N$.  If the adversary employs a collective attack (i.e., an attack where Eve attacks each round identically and independently) then, in the asymptotic scenario where $N \rightarrow \infty$, it was proven in \cite{QKD-Winter-Keyrate,QKD-renner-keyrate} that:
\begin{equation}\label{eq:keyrate}
r = \lim_{N\rightarrow \infty}\frac{\ell(N)}{N} = H(A|E)_\rho - H(A|B),
\end{equation}
where $\rho_{ABE}$ is the state modeling a single raw-key bit.  Namely, it describes Alice and Bob's raw key bit (as a random variable) and $E$'s quantum ancilla after the protocol is run, but \emph{before error correction and privacy amplification}.  Since collective attacks are iid (independent and identically distributed), the entire raw key can be described by the system $\rho_{ABE}^{\otimes N}$.  Often, one may promote security of collective attacks to general attacks and so collective attacks are usually analyzed in QKD security analyses \cite{QKD-survey1,QKD-survey2}.  We comment more on this later.

To compute the key-rate of a protocol using Equation \ref{eq:keyrate}, we therefore need to compute bounds on the von Neumann entropy.  To do so, later, we will use a theorem from \cite{QKD-Tom-Krawec-Arbitrary} which states:
\begin{theorem}\label{thm:entropy-bound}
  Let $\rho_{AE}$ be a classical-quantum state of the form:
  \[
  \rho_{AE} = \frac{1}{N}\kb{0}_A\otimes\left(\sum_{i=0}^m \kb{E_i}\right) + \frac{1}{N}\kb{1}_A\otimes\left(\sum_{i=0}^m \kb{F_i}\right).
  \]
  Then, it holds that:
  \[
  H(A|E)_\rho \ge \sum_{i=0}^m \left(\frac{\bk{E_i} + \bk{F_i}}{N}\right)\cdot\left(h\left[\frac{\bk{E_i}}{\bk{E_i}+\bk{F_i}}\right] - h[\lambda_i]\right),
  \]
  where:
  \[
  \lambda_i = \frac{1}{2}\left(1 + \frac{\sqrt{ (\bk{E_i} - \bk{F_i})^2 + 4Re^2\braket{E_i|F_i}}}{\bk{E_i} + \bk{F_i}}\right).
  \]
  Above, $h(x)$ is the binary Shannon entropy function, namely $h(x) = -x\log_2x - (1-x)\log_2(1-x)$.
\end{theorem}

\section{The Protocol}
The protocol we propose is an extension of the original M-SQKD protocol introduced in \cite{krawec2015mediated}.  That protocol discarded, in the best case, one half of all quantum signals sent by the server.  We modify this protocol to fully utilize all quantum signals, while also extending its capabilities to counter high noise channels.  Users Alice and Bob are ``semi-quantum'' and therefore restricted to performing the following actions each iteration:
\begin{enumerate}
  \item A user may choose to $\MR$ in which case the incoming signal is subjected to a $Z = \{\ket{0}, \ket{1}\}$ measurement resulting in $r\in\{0,1\}$. A qubit $\ket{r}$ is then sent back to the original sender.
  \item A user may choose to $\R$ in which case the incoming signal is reflected, without disturbance, back to the sender.
\end{enumerate}
Note that semi-quantum users are only able to communicate directly with the $Z$ basis or to ``disconnect'' from the quantum channel (in which case the sender, in our case the server, is ``talking to itself'').

\begin{figure}
    \centering
    \includegraphics[width=.9\textwidth]{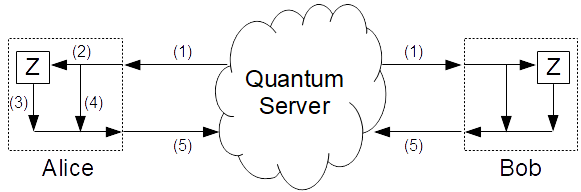}
    \caption{A diagram of our protocol.  A quantum server (which may be adversarial) prepares a quantum state (1) and sends part to Alice and part to Bob.  Next, Alice (and, independently, Bob) will choose $\MR$ or $\R$ (2).  If $\MR$, the signal for that round is subjected to a $Z$ basis measurement resulting in outcome $\ket{r}$ which also becomes the output of Alice's lab (3); otherwise, if $\R$ (4), then the incoming state is simply sent to the output of Alice's lab.  Similarly for Bob.  Finally, a signal returns to the quantum server (5) who is allowed to perform any operation on it, but must send a single classical message to Alice and Bob.  If the server is honest, the state prepared in (1) is the Bell state $\ket{\phi_0}$ while, on return (5), a Bell measurement is performed and the message is the actual Bell outcome received by the server.  We will prove security, however, assuming the server is adversarial and may not follow the protocol.}
    \label{fig:M-SQKD}
\end{figure}

The protocol acts as follows (a diagram can also be seen in Figure \ref{fig:M-SQKD}):

\begin{itemize}
\item \textbf{Quantum Communication Stage - } Repeat the below process until a sufficiently large raw key has been established.  A single round of this stage consists of:
  \begin{enumerate}
  \item The server, $C$, prepares the Bell state $\ket{\phi_0}$ and sends one qubit to Alice and one qubit to Bob.
  \item Alice and Bob choose, independently of one another, to $\MR$ or $\R$ their qubit.  We denote by $p_{M}$ to be the probability that a single party chooses $\MR$ and $p_R = 1-p_M$ to be the probability that a party chooses $\R$.
    \item The server, on receipt of both qubits, will perform a Bell measurement and announce the outcome as a classical message $\msg{0}$, $\cdots$, $\msg{3}$ to both Alice and Bob.
  \end{enumerate}
\item \textbf{Sampling Stage - } For every round in the above communication stage, Alice and Bob disclose their choice of $\MR$ or $\R$.  A subset $\tau$ of all rounds is chosen and all actions and measurement results (if applicable) are disclosed on those rounds in order to estimate the noise in the channel and the server's honesty (to be discussed).  Alice and Bob also disclose to one another the message they received from the server to ensure that both parties receive the same exact message for every round (both in and out of $\tau$).

  Based on the noise in the channel, Alice and Bob will also determine a setting for $\mode$ (either $\modeflip$ or $\modenoflip$).  In particular, users will estimate all needed probability values as used by our key-rate computation (discussed in Section \ref{section:key-rate}).  Once done, users may evaluate the expected key rate in the event users choose $\modeflip$ or $\modenoflip$ (we provide equations for both cases).  Once users determine which of the two provide a higher key-rate, users may set $\mode$ to that option and continue the protocol with that option (since the choice of $\mode$ only affects raw key generation below and not the completed quantum communication portion of the protocol users may choose it optimally, at this point).
\item \textbf{Raw Key Generation - } For every round not in $\tau$ and when both parties chose $\MR$, Alice and Bob will use their measurement outcomes as their raw key bits.  Furthermore, if $\modeflip$, then for every round where the server sent the message $\msg{2}$ or $\msg{3}$ (which should correspond to a Bell outcome of $\ket{\phi_2}$ or $\ket{\phi_3}$), Bob will flip his raw key bit; otherwise he leaves it alone.
\item \textbf{Postprocessing - } Alice and Bob will run an error correction protocol and privacy amplification to yield their final secret key.  See \cite{QKD-survey1} for details on these standard processes.
\end{itemize}

Note there are two major differences between our protocol and the original M-SQKD one from \cite{krawec2015mediated}.  First, raw key bits may be established regardless of the server's message whereas in \cite{krawec2015mediated}, raw key bits could only be distilled when the server sent the message $\msg{1}$.  Ideally, without noise and with an honest server, if both parties chose $\MR$, the server would send $\msg{1}$ only half of the time and, thus, half the signals were wasted originally.  This was originally done for security reasons, however we prove in this paper that security can be guaranteed even without this restriction, thus improving overall efficiency.  Furthermore, in the asymptotic scenario, $p_M$ may be set arbitrarily close to $1$ thus implying that every signal can be used for key distillation.  Thus, our protocol attains asymptotically perfect efficiency, unlike \cite{krawec2015mediated} which can only be at most $50\%$ efficient.

Secondly, our protocol allows for Bob to flip his raw key bit if the server sends the message $\msg{2}$ or $\msg{3}$.  This can be advantageous in some scenarios, as such a message implies that the server (if honest) received a Bell measurement of $\ket{\phi_2}$ or $\ket{\phi_3}$ indicating, for some noise scenarios, that there is a possibility that Alice and Bob's raw key bits are incorrect and so by flipping Bob's raw key bit, the correlation is restored.  Of course, care must be taken when using this option, as there are other noise scenarios where this flipping option can destroy the correlation thus creating more errors in the raw key.  The important observation, however, is that Alice and Bob may choose the setting of this value \emph{after} the Sampling Stage and so may use that data to determine an optimal strategy.  Later, we will evaluate our protocol in a variety of noise scenarios showing how this option can lead to drastic improvements in key generation rates.

\section{Security Analysis}\label{section:key-rate}

We first compute the key-rate of our protocol assuming collective attacks.  These are attacks where the adversary will perform the same (potentially probabilistic) attack each round of the quantum communication stage.  However, the adversary is also free to postpone measurement of her ancilla until any future point in time.  Later, we will show how this analysis may be promoted to security against general attacks where there are no restrictions on the adversary.  We will do this later by showing a novel reduction to a one-way entanglement based protocol; this reduction may hold applications to other M-SQKD protocols.

For our security proof, we assume the following:
\begin{enumerate}
\item Qubits are ideal and not subject to loss.  Furthermore, multi-qubit signals in a single round are not considered
\item Alice and Bob's devices are ideal.  We do not consider implementation level attacks such as photon tagging \cite{SQKD-photon-tag,SQKD-photon-tag-comment}.
\item The server $C$ may be controlled completely by the adversary.  Thus the server is allowed to send any signal state to Alice and Bob (subject to the above).  This signal may be entangled with a private ancilla held by $C$ (now the adversary) of arbitrary dimension.  On return of the two qubits from the users, the server may perform any quantum operation on these signals and the original ancilla state.  As a consequence of this, we do not need to consider third-party adversaries, as any such attack can be ``absorbed'' into an adversarial server's attack.  Also, the classical communication between the server and users does not need to be authenticated.  (Of course, the classical channel between Alice and Bob does need to be authenticated.)
\item The server must send a single classical message of $\msg{0}$, $\cdots$, $\msg{3}$ to parties on each round.  Furthermore, we assume the message sent to Alice and Bob is identical (that is, the adversary cannot send one message to Alice and a different message to Bob on a single round).  This assumption is easy to enforce given that Alice and Bob reveal to each other, using their authenticated channel, all messages received by the server (both those within and without the sample subset $\tau$).  Users may then abort if a single message is different between users for a particular round.
\end{enumerate}

Assumptions (1) and (2) are useful for determining the theoretical performance of our system and also for comparing to other SQKD protocols for which most make these two assumptions.  These assumptions may be dropped, or loosened, perhaps using the techniques in \cite{boyer2017experimentally,massa2019experimental} combined with our security proof below.  However, we leave that analysis as interesting future work.

\subsection{Key Rate Derivation}

An adversarial server may prepare any state it likes in step 1 of the quantum communication stage.  In particular, the server may prepare and send the state
\[\ket{\widetilde{\psi}_0} = \sum_{i,j\in\{0,1\}}\beta_{i,j}\ket{i,j}_{AB}\otimes\ket{E_{i,j}}_E.\]
The $A$ and $B$ qubits are sent to Alice and Bob respectively while Eve (the server) will keep the $E$ portion.

However, it was proven in \cite{krawec2019multi} (see Theorem 1 of that source), that for a protocol of this form, it is sufficient to actually consider an initial state of the form
\[\ket{\psi_0} = \sum_{i,j \in \{0,1\}}\alpha_{i,j}\ket{i,j}\]
which is not entangled with Eve's ancilla and where, furthermore, each $\alpha_{i,j}$ is real and non-negative.  Thus, we only need to prove security assuming the server sends the ``simpler'' state $\ket{\psi_0}$ and by Theorem 1 of \cite{krawec2019multi}, security against arbitrary initial states of the form $\ket{\widetilde{\psi}_0}$ will follow.

In the return channel, when qubits return to the server (Eve), she is allowed to perform any quantum operation of her choice, potentially creating an entanglement with a private ancilla.  However, the server must send a message to both Alice and Bob and, as discussed, the message must be identical for both parties.  This entire process can be modeled as a quantum instrument \cite{davies1970operational}.  Furthermore, using standard techniques \cite{wilde2011classical}, this quantum instrument may actually be dilated to an isometry (which may then be extended to a unitary operator).  In particular, the attack will consist of an operator $U$, mapping the two returning qubits to Eve's private ancilla and a four dimensional Hilbert space $\mathcal{H}_{cl}$ spanned by $\{\ket{\msg{0}}, \cdots, \ket{\msg{3}}\}$.  The attack will consist of Eve applying this operator and then performing a projective measurement on the $\mathcal{H}_{cl}$ space in this particular basis.  The measurement outcome determines the message she sends to parties and the post measurement state of the ancilla and qubit system determines the state of her private ancilla in the event she had sent that message using a quantum instrument.  For details, along with an explicit proof that this is equivalent to a quantum instrument attack against M-SQKD protocols, the reader is referred to \cite{krawec2015mediated}.  Without loss of generality, this isometry may be described by its action on basis states as follows:
\begin{equation}\label{eq:attack-op}
    U\ket{i,j} = \sum_{m=0}^3\ket{\msg{m}, e_{i,j}^m}_{cl,E}.
\end{equation}
Note that, above, the $\ket{e_{i,j}^m}$ are arbitrary, and not necessarily normalized, states in Eve's ancilla.

We are now ready to derive a bound on the \-asymptotic key-rate of our protocol against collective attacks.  To do so, we must derive a density operator describing the joint state of the Alice, Bob, and Eve systems conditioning on a raw-key being distilled.  This will allow us to compute the required entropies needed to compute the key-rate of the protocol using Equation \ref{eq:keyrate}.  In this case, the server sends the initial state $\ket{\psi_0}$ as discussed while Alice and Bob choose $\MR$ (as we are conditioning on events that lead to a raw key bit for this round).  When the two qubits return to the server, the adversary will apply $U$ and measure the ``cl'' message subspace.  This leads to the following mixed state:
\begin{equation}\label{eq:state-no-flip}
    \rho_{ABE} = \sum_{i,j\in\{0,1\}}\alpha_{i,j}^2 \kb{i,j}_{AB} \otimes \sum_{m=0}^3\kb{\msg{m}, e_{i,j}^m}.
\end{equation}

Now, if $\modenoflip$, then $\rho_{ABE}$ is the final joint state.  If $\modeflip$, then Bob will flip his raw key bit in the event he receives message $\msg{2}$ or $\msg{3}$.  In this case, the state of the system becomes:
\begin{align}
    \sigma_{ABE} &= \sum_{i,j\in\{0,1\}}\alpha_{i,j}^2 \kb{i,j}_{AB} \otimes \sum_{m=0}^1\kb{\msg{m}, e_{i,j}^m}\label{eq:state-flip}\\
    &+ \sum_{i,j\in\{0,1\}}\alpha_{i,j}^2 \kb{i,1-j}_{AB} \otimes \sum_{m=2}^3\kb{\msg{m}, e_{i,j}^m}.\notag
\end{align}
However, observe that $tr_B\rho_{ABE} = tr_B\sigma_{ABE}$. In particular:
\begin{align*}
  \rho_{AE} = \sigma_{AE} &= \kb{0}_A\otimes\sum_j\sum_m\alpha_{0,j}^2\kb{\msg{m},e_{0,j}^m}\\
  &+ \kb{1}_A\otimes\sum_j\sum_m\alpha_{1,j}^2\kb{\msg{m},e_{1,j}^m}.
\end{align*}
Thus, it suffices to bound $H(A|E)_\rho$ for both choices of $\mode$.
Note that, even though $\mode$ does not affect $H(A|E)$, it will affect $H(A|B)$ and thus will play an important part later in our analysis.

Our goal now is to compute a bound on $H(A|E)_\rho$ which will give us also the entropy needed for the case of $\sigma_{AE}$.  To do so, we use Theorem \ref{thm:entropy-bound} which, applied to the above state, yields the following result:
\begin{equation}\label{eq:entropy}
H(A|E)_\rho \ge \sum_{j,m}\left(\alpha_{0,j}^2\bk{e_{0,j}^m} + \alpha_{1,1-j}^2\bk{e_{1,1-j}^m}\right)H_{j,m},
\end{equation}
where:
\begin{equation}
  H_{j,m} = h\left(\frac{\alpha_{0,j}^2\bk{e_{0,j}^m}}{\alpha_{0,j}^2\bk{e_{0,j}^m} + \alpha_{1,1-j}^2\bk{e_{1,1-j}^m}}\right) - h\left(\lambda_{j,m}\right),
\end{equation}
and finally:
\begin{equation}\label{eq:lambda}
  \lambda_{j,m} = \frac{1}{2}\left( 1 + \frac{\sqrt{(\alpha_{0,j}^2\bk{e_{0,j}^m} - \alpha_{1,1-j}^2\bk{e_{1,1-j}^m})^2 + 4\alpha^2_{0,j}\alpha^2_{1,1-j}Re^2\braket{e_{0,j}^m|e_{1,1-j}^m}}}{\alpha_{0,j}^2\bk{e_{0,j}^m} + \alpha_{1,1-j}^2\bk{e_{1,1-j}^m}}\right)
\end{equation}

The reader will note that we applied Theorem \ref{thm:entropy-bound} by setting the $\ket{E_i}$ terms to be of the form $\alpha_{0,j}\ket{e_{0,j}^m}$ and the corresponding vectors $\ket{F_i}$ of the form $\alpha_{1,1-j}\ket{e_{1,1-j}^m}$.  To use Theorem \ref{thm:entropy-bound}, one requires a ``pairing'' of Eve's vectors in the even Alice has a key-bit of zero with that of a key-bit of one.  Any pairing provides a lower-bound on the entropy, however care must be taken to choose a pairing that provides the most optimistic result.  In general, due to the way in which Theorem \ref{thm:entropy-bound} is proven (see \cite{QKD-Tom-Krawec-Arbitrary}), it is best to pair similar events with each other.  Thus, we pair, for instance, $\alpha_{0,0}\ket{e_{0,0}^m}$ with $\alpha_{1,1}\ket{e_{1,1}^m}$ as they both represent the event that there is no error in the raw key and the same message ``$m$'' was sent.  Other pairings, such as $\alpha_{0,0}\ket{e_{0,0}^m}$ with, say, $\alpha_{1,0}\ket{e_{1,0}^m}$, though providing us with a lower-bound that is easier to compute than the one we derive, actually produces a substantially lower entropy bound.  Furthermore, pairing vectors with different message outcomes (e.g., $m$ and $m'$) produces a worse bound as it is impossible for Alice and Bob to determine information on the overlap of Eve's vectors in this case based only on observed parameters.

Thus, to evaluate the von Neumann entropy, needed to compute the key-rate of our protocol (Equation \ref{eq:keyrate}), we must now find bounds on the inner-products and $\alpha$ values appearing in the above expressions.  These bounds, however, must be functions only of observable parameters which Alice and Bob can directly determine.

We begin by defining some notation.  For $i,j \in \{0,1\}$, let $P_{i,j}$ be the probability that Alice observes $\ket{i}$ and Bob observes $\ket{j}$ conditioning on them both choosing $\MR$.  Clearly:
\begin{equation}\label{eq:alpha}
  P_{i,j} = \alpha_{i,j}^2.
\end{equation}

Next, let $P_{i,j}^m$, for $i,j\in\{0,1,R\}$, be the probability that the server sends message $\msg{m}$, conditioning on Alice choosing $\MR$ and observing $\ket{i}$ (if $i = 0,1$) or Alice choosing $\R$ (if $i=R$) and similarly for Bob with $j$.  It is not difficult to see that
\begin{equation}\label{eq:message-pr}
P_{i,j}^m = \bk{e_{i,j}^m} \text{ when $i,j\in\{0,1\}$}.
\end{equation}

In the next section, we will consider values of the form $P_{R,j}^m$ and $P_{i,R}^m$ which turn out to be very important in bounding Eve's information for our protocol.  Values of this form may be considered a form of mismatched measurement, introduced originally in \cite{QKD-Tom-First} for standard QKD analysis, and have been used extensively lately to boost noise tolerance of various (S)QKD protocols \cite{QKD-Tom-KeyRateIncrease,QKD-Tom-KeyRateMismatchedDistill,QKD-Tom-Krawec-Arbitrary}.  We will also require $P^m_{R,R}$ for our entropy bound.

\subsubsection{Mismatched Events}

Let us first consider $P_{R,0}^m$, namely, the probability that the server sends message $\msg{m}$, conditioning on Alice choosing $\R$ and Bob choosing $\MR$ and observing $\ket{0}$.  To determine this value, we trace the evolution of the protocol.  Initially, the server sends $\ket{\psi_0} = \sum_{i,j}\alpha_{i,j}\ket{i,j}$.  Conditioning on Bob observing $\ket{0}$ and Alice ignoring her system, the state collapses to:
\[
\frac{\alpha_{0,0}\ket{0,0} + \alpha_{1,0}\ket{1,0}}{\sqrt{\alpha_{0,0}^2 + \alpha_{1,0}^2}}.
\]
When this state returns to the server, she will attack with operator $U$ defined in Equation \ref{eq:attack-op}, evolving the system to
\[
\sum_{m=0}^3\ket{\msg{m}}\otimes\left(\frac{\alpha_{0,0}\ket{e_{0,0}^m} + \alpha_{1,0}\ket{e_{1,0}^m}}{\sqrt{\alpha_{0,0}^2+\alpha_{1,0}^2}}\right).
\]
From this, we attain the desired probability value:
\begin{align}
  P_{R,0}^m &= \frac{\alpha_{0,0}^2\bk{e_{0,0}^m} + \alpha_{1,0}^2\bk{e_{1,0}^m} + 2\alpha_{0,0}\alpha_{1,0}Re\braket{e_{0,0}^m|e_{1,0}^m}}{\alpha_{0,0}^2 + \alpha_{1,0}^2}\notag\\
  &=\frac{P_{0,0}P_{0,0}^m + P_{1,0}P_{1,0}^m + R_{0010}^m}{P_{0,0} + P_{1,0}}\notag
\end{align}
where, above, we used Equations \ref{eq:alpha} and \ref{eq:message-pr} and we also define:
\begin{equation}
  R_{xyzw}^m = 2\sqrt{P_{x,y}P_{z,w}}Re\braket{e_{x,y}^m|e_{z,w}^m}.
\end{equation}

Through a similar process, we may compute the following values:
\begin{align}
  P_{i,R}^m &= \frac{P_{i,0}P_{i,0}^m + P_{i,1}P_{i,1}^m + R_{i0i1}^m}{P_{i,0} + P_{i,1}}\\\notag\\
  P_{R,j}^m &= \frac{P_{0,j}P_{0,j}^m + P_{1,j}P_{1,j}^m + R_{0j1j}^m}{P_{0,j} + P_{1,j}}
\end{align}

Critically, the above analysis allows us to learn the exact value of important inner products of the form $Re\braket{e_{i,0}^m|e_{i,1}^m}$ and $Re\braket{e_{0,j}^m|e_{1,j}^m}$.  In particular:
\begin{align}
  R_{i0i1}^m &= (P_{i,0} + P_{i,1})P^m_{i,R} - P_{i,0}P_{i,0}^m - P_{i,1}P_{i,1}^m\\\notag\\
  R_{0j1j}^m &= (P_{0,j} + P_{1,j})P^m_{R,j} - P_{0,j}P_{0,j}^m - P_{1,j}P_{1,j}^m.
\end{align}
Note that the right hand side of both expressions above involve only observable statistics which Alice and Bob can estimate in the Sampling stage of the protocol.  These will be important momentarily.

\subsubsection{Reflection Error Events}

The final important statistic which Alice and Bob must consider is the probability that the server sends a particular message $\msg{m}$ conditioning on both parties choosing $\R$.  We denote this value by $P_{R,R}^m$.  Note that, ideally, this message should always be $\msg{0}$ and, so, any alternative message sent in this case can be considered an error, either due to a malicious server, phase error in the channel, or both.  Note, we do not distinguish between natural noise and adversarial noise and simply assume the worst case that all errors in the signal or messaging is due to an adversarial attack.  It turns out that this expression, combined with the above, will yield critical information on the inner products appearing in Equation \ref{eq:lambda}. In particular, we need information on $R_{0011}^m$ and $R_{0110}^m$.

To determine $P_{R,R}^m$ as a function of the inner products of Eve's ancilla (which will give us the necessary information to evaluate Equation \ref{eq:entropy}), we again trace the evolution of the protocol, now conditioning on both parties choosing $\R$.  In this case, the server sends $\ket{\psi_0}$ and both parties ignore the signal, reflecting it back.  Since we are assuming all noise in the channel is adversarial (i.e., all noise is the result of an adversary), the state, therefore arrives back at the server in this form.  The server then applies $U$ which evolves the system to:
\[
\sum_{m=0}^3\ket{\msg{m}}\otimes\left(\sum_{i,j\in\{0,1\}}\alpha_{i,j}\ket{e_{i,j}^m}\right).
\]
From this, it is easy to show that:
\begin{align*}
  P_{R,R}^m = \sum_{i,j}\alpha_{i,j}^2\bk{e_{i,j}^m} + R_{0001}^m + R_{0010}^m + R_{0011}^m + R_{0110}^m + R_{0111}^m + R_{1011}^m
\end{align*}

Using the above analysis, this implies:
\begin{align}
  R_{0011}^m + R_{0110}^m = P_{R,R}^m - \sum_{i,j}P_{i,j}P_{i,j}^m &- [(P_{0,0} + P_{0,1})P^m_{0,R} - P_{0,0}P_{0,0}^m - P_{0,1}P_{0,1}^m]\label{eq:R0011R0110}\\
  &-[(P_{0,0} + P_{1,0})P^m_{R,0} - P_{0,0}P_{0,0}^m - P_{1,0}P_{1,0}^m]\notag\\
  &-[(P_{0,1} + P_{1,1})P^m_{R,1} - P_{0,1}P_{0,1}^m - P_{1,1}P_{1,1}^m]\notag\\
  &-[(P_{1,0} + P_{1,1})P^m_{1,R} - P_{1,0}P_{1,0}^m - P_{1,1}P_{1,1}^m]\notag
\end{align}

\subsection{Final Key-Rate Bound}
This gives us everything we need to compute the von Neumann entropy in Equation \ref{eq:entropy}.  In particular, we minimize Equation \ref{eq:entropy} subject to the above constraints, all of which are functions of observable statistics.  Additionally, the following constraints may be derived through use of the Cauchy-Schwarz inequality:
\begin{align*}
    |R_{0011}^m| &\le 2\sqrt{P_{0,0}P_{1,1}}\sqrt{\bk{e_{00}^m}\bk{e_{11}^m}} = 2\sqrt{P_{0,0}P_{1,1}}\sqrt{P_{0,0}^m\cdot P_{1,1}^m}\\
    |R_{0110}^m| &\le 2\sqrt{P_{0,1}P_{1,0}}\sqrt{\bk{e_{01}^m}\bk{e_{10}^m}} = 2\sqrt{P_{0,1}P_{1,0}}\sqrt{P_{0,1}^m\cdot P_{1,0}^m}
\end{align*}

To perform the minimization, we note that the function to be minimized, namely the right hand side of Equation \ref{eq:entropy} is convex and we are optimizing over a closed interval.  Indeed, first note that the free parameters to optimize over are $\{R_{0110}^m\}_{m=0}^3$.  This is due to the fact that $R_{0011}^m$ is a function of $R_{0110}^m$ and some constant value (that constant being a simple function of the observed probability values as shown in Equation \ref{eq:R0011R0110}).  Next, note that the right hand side of Equation \ref{eq:entropy} which we are minimizing, can be broken up into four independent functions of the form $f_m(R_{0110}^m)$ which is the sum of the terms involving $H_{0,m}$ and $H_{1,m}$.  Thus, we may minimize each $f_m$ separately.  Finally, note that the function $f_m(\cdot)$ is of the form:
\begin{equation}
c_{0,m}\left(d_{0,m} - h\left[\frac{1}{2} + \frac{\sqrt{e_{0,m} + 4(R_{0110}^m)^2}}{h_{0,m}}\right]\right) + c_{1,m}\left(d_{1,m} - h\left[\frac{1}{2} + \frac{\sqrt{e_{1,m} + 4(g_{1,m}-R_{0110}^m)^2}}{h_{1,m}}\right]\right)
\end{equation}
where $c_{i,m}, d_{i,m}, e_{i,m}, h_{i,m},$ and $g_{i,m}$ are constants (functions of the observed probability values) and $c_{i,m}$, $d_{i,m}$, $e_{i,m}$, and $h_{i,m}$ are positive.  We claim this is a continuous convex function in the parameter to be optimized thus the minimum exists.  Indeed, the function $r(x) = -h(1/2 + \sqrt{a + 4x^2}/b)$ is convex for positive $a$ and $b$.  To see this, note that $h(1/2+x)$ is concave and non increasing for $x \in [0,1/2]$ and that $\sqrt{a+4x^2}/b$ is convex (for positive $a$ and $b$); thus their composition is concave and so its negative is convex.  Thus both the first term in the above is convex and so is the second as that results from the composition with an affine transformation.  To actually evaluate the minimum in the subsequent section, we used Mathematica's \texttt{NMinimize} function.

The only remaining element to compute is the conditional Shannon entropy $H(A|B)$.  However, this is easy to compute given observed statistics.  Indeed, it is a function only of the probability distribution of Alice and Bob's raw key bits.  Let $P^{key}_{a,b}$ be the probability that Alice's raw key bit is ``$a$'' and Bob's raw key bit is ``$b$.''  It is not difficult to see from Equations \ref{eq:state-no-flip} and \ref{eq:state-flip} that these are, for $\modenoflip$:
\begin{align}
    P^{key}_{i,j} = P_{i,j}\sum_m P_{i,j}^m
\end{align}
and when $\modeflip$ we have:
\begin{align}
    P^{key}_{i,j} = P_{i,j}\left(P_{i,j}^0 + P_{i,j}^1\right) + P_{i,1-j}\left(P_{i,1-j}^2 + P_{i,1-j}^3\right)
\end{align}

This allows us to readily compute:
\[
H(A|B) = H\left(P^{key}_{0,0}, \cdots, P^{key}_{1,1}\right) - h\left(P^{key}_{0,0} + P^{key}_{1,0}\right)
\]
thus completing the key-rate derivation.

\subsection{Evaluation}

Our security proof above works for any noise signature.  That is, one simply needs to observe all ``$P$'' values appearing in the above expressions and analysis and perform the minimization of $H(A|E)$.  However, to actually evaluate our key-rate bound, we will assume a depolarization channel.  This is a common assumption in QKD security proofs and also allows us to compare with prior work.  We will also, in order to determine $P$ values to evaluate, assume the server follows the protocol honestly. \emph{Note that none of this is a requirement of the proof,} it is simply a way to evaluate our bound as we must put numbers to those statistics appearing in our key-rate derivation.

A depolarization channel with parameter $Q$ takes a two qubit quantum state $\rho$ and maps it to:
\[
\mathcal{E}_Q(\rho) = (1-2Q)\rho + Q\frac{I}{2}.
\]
(Here, $I$ is the dimension four identity operator.)
We choose this particular parameterization so that $Q$ becomes more directly related with the error in Alice and Bob's measurements as will soon be evident.  We will assume independent depolarization channels in the forward and reverse channel, using $Q_F$ to denote the depolarization parameter in the Forward channel (from the server to Alice and Bob, and $Q_R$ to denote the parameter in the Reverse channel (from Alice and Bob to the server).

Using this, we may parameterize the many noise statistics needed for our key-rate computation.  These are easily seen to be:
\begin{align*}
    \mathrm{P}_{0,0} &= P_{1,1} = \frac{1}{2}(1-Q_F)\\
    \mathrm{P}_{0,1} &= P_{1,0} = \frac{1}{2}Q_F
\end{align*}
\begin{align*}
    \mathrm{P}_{00}^0 &= \mathrm{P}_{00}^1 = \mathrm{P}_{11}^0 = \mathrm{P}_{11}^1 = \frac{1}{2}(1-Q_R)\\
    \mathrm{P}_{00}^2 &= \mathrm{P}_{00}^3 = \mathrm{P}_{11}^2 = \mathrm{P}_{11}^3 = \frac{1}{2}Q_R\\
    \mathrm{P}_{01}^0 &= \mathrm{P}_{01}^1 = \mathrm{P}_{10}^0 = \mathrm{P}_{10}^1 = \frac{1}{2}Q_R\\
    \mathrm{P}_{01}^2 &= \mathrm{P}_{01}^3 = \mathrm{P}_{10}^2 = \mathrm{P}_{10}^3 = \frac{1}{2}(1-Q_R).\notag
\end{align*}

For the reflection events, the system passes through both channels sequentially.  Thus, to model the case when both Alice and Bob reflect, we use $\mathcal{E}_{Q_R}(\mathcal{E}_{Q_F}(\kb{\phi_0}))$ to derive:
\begin{align*}
    \mathrm{P}_{RR}^0 &= (1-2Q_R)(1-2Q_F) + \frac{1}{2}(1-2Q_R)Q_F + \frac{1}{2}Q_R\\
    \mathrm{P}_{RR}^1 &= P_{RR}^2 = P_{RR}^3 =  \frac{1}{2}(1-2Q_R)Q_F + \frac{1}{2}Q_R
\end{align*}

Finally, we need values for $P_{j,R}^m$ and $P_{R,j}^m$.  To do this, we first apply the depolarization channel in the forward direction and then simulate Alice and Bob's measurements, conditioning on the required outcome.  From this post measured state, we apply the depolarization channel again (for the return trip to the server) and calculate the desired probabilities.  This process leads us to the following derivations:
\begin{align*}
    P_{R,j}^0 = P_{j,R}^0 &= \frac{1}{2}[(1-2Q_R)(1-2Q_F) + (1-2Q_R)Q_F + Q_R]\\
    P_{R,j}^1 = P_{j,R}^1 &= \frac{1}{2}[(1-2Q_R)(1-2Q_F) + (1-2Q_R)Q_F + Q_R]\\
    P_{R,j}^2 = P_{j,R}^2 &= \frac{1}{2}[(1-2Q_R)Q_F + Q_R]\\
    P_{R,j}^3 = P_{j,R}^3 &= \frac{1}{2}[(1-2Q_R)Q_F + Q_R]\\
\end{align*}

We compare the key-rate of our protocol to the original M-SQKD protocol from \cite{krawec2015mediated}.  To perform this comparison, we use improved key-rate results from \cite{sqkd-med-improved}.  Note that the protocol of \cite{massa2019experimental} will always be less efficient due to its design choices (that protocol was designed to operate with practical devices whereas ours here is a more theoretical construction - while it may potentially be made practical using techniques from \cite{boyer2017experimentally,massa2019experimental}, this will lower its efficiency and so we do not compare these).  We also cannot compare to other M-SQKD protocols, (even those with asymptotically perfect efficiency \cite{m-2-liu2018mediated,m-3-lin2019mediated,m-9-chen2021efficient,m-10-hwang2020mediated} as ours has) as no other M-SQKD protocols have information theoretic key-rate derivations and so there is no statistic to compare (we can only compare the noiseless case in which case all these protocols, and ours, have full efficiency).  Note that, as discussed earlier, our proof methods may be applicable to those other protocols; though, of course, performing the necessary key rate computations for these alternative protocols is outside the scope of this work.

\begin{figure}
    \centering
    \includegraphics[width=.7\textwidth]{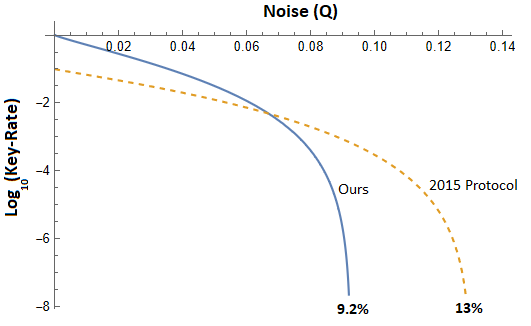}
    \caption{Evaluating the key-rate bound of our new protocol here (solid line) and comparing with the key-rate of the original M-SQKD protocol from 2015 \cite{krawec2015mediated}.  Here we have $Q_F = Q_R = Q$ and so there is no difference between the two settings for $\mode$.  We observe that the noise tolerance of our new protocol is lower, though the efficiency can be substantially higher for lower noise levels (lower than $6.5\%$ in this instance).  Since our protocol is ``backwards compatible'' one may actually use our protocol for lower levels of noise and switch to the original (normally less efficient) protocol from \cite{krawec2015mediated} if the observed channel noise is high enough to make the switch worthwhile.  Since the difference in our protocol and the original 2015 M-SQKD one is purely in the classical stage, this decision may be made after determining the channel noise and so an optimal choice may always be made.}
    \label{fig:1}
\end{figure}

\begin{figure}
    \centering
    \includegraphics[width=.7\textwidth]{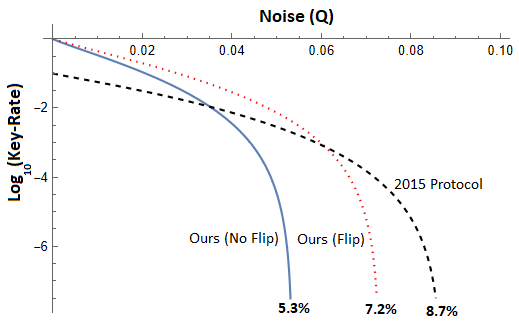}
    \caption{Evaluating our key-rate bound and comparing to the original 2015 M-SQKD protocol from \cite{krawec2015mediated}.  Here we set $Q_F = 2Q$ and $Q_R = Q$ (thus there is twice as much noise in the forward channel from the server to Alice and Bob as in the reverse channel.  In this asymmetric case, the choice of $\mode$ is important and can improve performance.  Again, we observe that our new protocol is more efficient except at higher noise levels. Note that $\mode$ may be decided on after running the sampling stage so that there is never a ``wrong choice.''}
    \label{fig:2}
\end{figure}

\begin{figure}
    \centering
    \includegraphics[width=.7\textwidth]{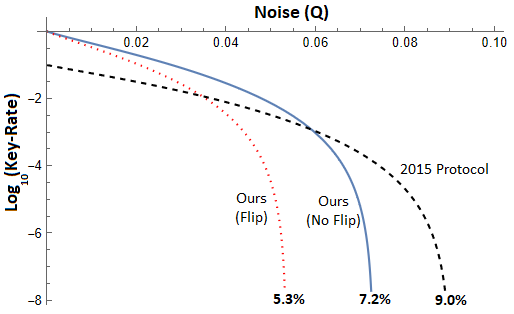}
    \caption{Similar to Figure \ref{fig:2} but now setting $Q_F = Q$ and $Q_R = 2Q$ (that is, twice as much noise in the reverse channel).}
    \label{fig:3}
\end{figure}

The evaluation and comparison is shown in Figures \ref{fig:1}, \ref{fig:2}, and \ref{fig:3}.  Figure \ref{fig:1} shows the case when $Q_R = Q_F = Q$.  Note that in this case there is no difference in $\modeflip$ and $\modenoflip$ which is clear from the protocol construction.  Figure \ref{fig:2} shows the case when $Q_F = 2Q$ and $Q_R = Q$ (namely, the noise in the forward channel is twice as high as the noise in the reverse channel).  Finally, Figure \ref{fig:3} shows the case when $Q_F = Q$ and $Q_R = 2Q$ (i.e., the noise in the reverse channel is twice the noise in the forward).

In all cases we note that our new protocol greatly outperforms the original M-SQKD protocol on which it was based when the noise level is not too large.  As the noise increases, our new protocol will reach a zero key-rate before the original does.  This is not surprising.  Our new protocol utilizes all messages from the server and so efficiency naturally increases, as is observed for smaller noise levels.  As the noise level increases, however, and Eve (the server) potentially gathers more information on Alice and Bob's raw key, the fact that the original protocol only used cases when the server sent a single specific message (translating normally to a Bell measurement of $\ket{\phi_1}$ only), actually aids Alice and Bob by decreasing efficiency but increasing noise tolerance.  Namely, there are more attack strategies available to a server who is allowed to use all four messages whereas in the original, only a single message led to a key bit being distilled.

Finally, we note that, as expected, the use of $\modeflip$ and $\modenoflip$ is important depending on the relative noise levels of the forward and reverse channels.  However, as mentioned before, the choice of $\mode$ may be made after channel statistics are gathered (as it is purely a classical operation on the raw key data).  Thus, Alice and Bob may run the quantum portion of the protocol, estimate the channel noise, and then decide on an optimal choice for the $\mode$.  Pushing this idea further, Alice and Bob may even decide whether or not to use the original M-SQKD protocol from \cite{krawec2015mediated} or our new extension, as the new protocol we described here is backwards compatible.  Thus, taken as a whole, our work in this paper has shown how greatly improved efficiency is possible while still maintaining the high noise tolerance of the original M-SQKD protocol.

\subsection{Extension to General Attacks}

The previous section analyzed the security of our protocol assuming collective attacks.  However, we can extend this to security against general attacks by first showing an equivalent entanglement based protocol and then using de Finetti \cite{konig2005finetti} or postselection \cite{christandl2009postselection} techniques to promote our earlier analysis to the general case \cite{QKD-survey1}.  Note that this reduction to an entanglement based protocol is perhaps our largest contribution in this work as, prior to this, no reduction for M-SQKD protocols was known (only reductions for some classes of two-party SQKD protocols were constructed in \cite{krawec2018key,iqbal2020high}, however they did not apply to mediated SQKD protocols).  Such a reduction is important to proving security against general attacks using de Finetti style arguments \cite{privateconversation}.  Thus, our work in this section may be beneficial to other protocol security analysis, both in semi-quantum and general multi-user QKD scenarios.

In the prepare-and-measure scenario considered before, the server sent a quantum state to Alice and Bob who then returned a quantum state back to the server (and, of course, this server may be adversarial).  Instead, we will show this is equivalent to a scenario where an adversary prepares a quantum state, sending part of it to Alice, part to Bob, and part to a trusted server $C$, while also holding a part $E$ in a private ancilla.  Note that in the entanglement based version, the server is honest (though the source is not), however security in this setting will imply security in the ``real'' prepare-and-measure case even when the server is adversarial as we will show.

The entanglement based protocol operates as follows:
\begin{itemize}
    \item \textbf{Quantum Communication Stage:}
    \begin{enumerate}
        \item A quantum source (potentially adversarial) prepares a quantum state $\ket{\psi}_{ABCE}$ where the $A$ and $B$ registers consist of $N$ qubits each and the $C$ register consists of $N$ qudits, each of dimension $4$ (thus, $C$'s register is of total dimension $4^N$).  The $E$ system is kept private by the adversary and its dimension is arbitrary.
        \item Alice and Bob choose, independently and for each of the $N$ signals, $\MR$ or $\R$ (though, we note, these labels no longer have direct meaning in this case as Alice and Bob will not reflect anything).  If the choice is $\MR$, that party will measure their qubits in the $Z$ basis; otherwise, they will measure in the $X$ basis \emph{and abort if they observe $\ket{-}$}.
        \item The trusted user $C$ will measure his system in the computational basis $\{\ket{0}, \cdots, \ket{3}\}$ and report the outcome publicly.
        \item Alice and Bob will disclose their choices of $\MR$ and $\R$.  They will also disclose a random subset of their outcomes for quantum sampling purposes.  For those that were not disclosed, and for which both parties chose $\MR$, they will keep that round to contribute towards their raw key.  If the server sends the message $\msg{2}$ or $\msg{3}$ and $\modeflip$, Bob will flip his raw key bit for that round.
    \end{enumerate}
    \item \textbf{Postprocessing - } Same as in the prepare-and-measure protocol.
\end{itemize}

\begin{figure}
    \centering
    \includegraphics[width=.9\textwidth]{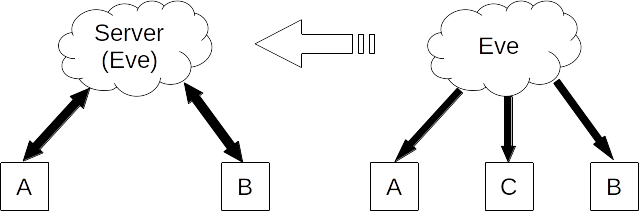}
    \caption{We show in this section how security of the entanglement based, fully quantum, protocol (Right) implies security of the semi-quantum prepare and measure protocol (left).  For the M-SQKD protocol, the adversarial server prepares and later receives quantum states from Alice and Bob and must report a classical message.  In the entanglement protocol, a quantum source (potentially adversarial) prepares a quantum state sending part to Alice, part to Bob, and part to a trusted server $C$.  We show that for any general attack against the SQKD protocol there exists an attack against the entanglement based protocol which creates an identical quantum state following the successful completion of the protocol.  Thus, security of the entanglement based protocol will imply security of the M-SQKD protocol as there can only be more attacks against the entanglement based version.}
    \label{fig:M-SQKD-proof}
\end{figure}

Note that this protocol is not a semi-quantum one.  Indeed, past reductions of semi-quantum to one-way fully quantum protocols involve the reduction to a particular fully quantum QKD protocol \cite{krawec2018key,iqbal2020high} and this is the same in the mediated case here.  The second interesting point is that users will abort if they ever observe a $\ket{-}$.  Thus, this entanglement based protocol is highly inefficient; however it is only a ``toy'' protocol and not meant to actually be used.  Instead, we will prove that, conditioning on a non-abort, its security implies the security of the prepare-and-measure protocol (where users are semi-quantum and do not have this abort case).  Further, we will show, that our previous analysis can be applied to the security of this entanglement based protocol.  Ultimately, our goal in this section is to prove the following security relations:
\[
\texttt{Col-PM} \Longrightarrow \texttt{Col-Ent} \Longrightarrow \texttt{Gen-Ent} \Longrightarrow \texttt{Gen-PM},
\]
where $X \Longrightarrow Y$ means that security of protocol ``$X$'' implies security of protocol ``$Y$'' and, furthermore, conditioning on protocols $X$ and $Y$ not aborting, the key rate of $Y$ will be no less than the key-rate of $X$ under the same channel noise conditions.  Above, we use ``\texttt{Col}'' to mean the protocol under collective attacks (``\texttt{Gen}'' for general attacks); we use ``\texttt{PM}'' to denote the prepare-and-measure SQKD protocol and ``\texttt{Ent}'' to denote the entanglement based protocol introduced in this section.  Standard techniques will be used for showing $\texttt{Col-Ent} \Longrightarrow \texttt{Gen-Ent}$, while we already showed that $\texttt{Col-PM}$ is secure in the previous section. Thus, our primary work will be to show the other relations.  See also Figure \ref{fig:M-SQKD-proof}.

\subsubsection{Non-Interactive Attacks}
We first show that security of the entanglement based protocol will imply security of the prepare and measure protocol for general, non interactive, attacks (i.e., those where the adversary can create any arbitrary quantum state, but must send all qubits to Alice and Bob simultaneously).  Later, we will show the more difficult case of interactive attacks where the adversary is allowed to adapt its strategy after receiving qubit(s) back from Alice and/or Bob.

For the reduction, we must show two things. First, we show that for any general attack (i.e., not necessarily a collective attack) against the prepare-and-measure protocol, there exists an equivalent attack in the entanglement based protocol.  Here, by ``equivalent'' we mean, conditioning on the entanglement protocol not aborting, the resulting quantum systems are identical in both the entanglement and the prepare and measure protocol and, so, any entropy computation for one will follow to the other.  Second, for any attack against the prepare and measure protocol, the equivalent attack produces a system with a non-zero probability of not aborting (otherwise, we would be conditioning on a probability zero event).  Note that there are attacks against the entanglement version which will always cause it to abort (e.g., the source Eve can simply send all $\ket{-}$'s to Alice and Bob) however such states are impossible to appear in the prepare and measure protocol and so are not worth considering.  Taken together, this will show $\texttt{Gen-Ent} \Longrightarrow \texttt{Gen-PM}$ (as the entanglement based version can only have more attacks against it).

For the prepare and measure protocol, a general attack will be modeled as an adversarial server preparing an arbitrary $2N$ qubit state, entangled with its ancilla.  Here, $N$ will be the number of rounds in the protocol.  Half the qubits are sent to Alice and the other half to Bob.  After Alice and Bob perform their operations on their respective qubits, $2N$ qubits return to the server who is allowed to perform an arbitrary attack on all qubits simultaneously using a quantum instrument which may also act on the server's initial, private, quantum ancilla.  From this, $N$ classical messages are sent to both Alice and Bob.  For the entanglement based protocol, a general attack will consist of Eve preparing any arbitrary state, sending $N$ qubits to Alice, $N$ qubits to Bob, and $N$ dimension four qudits to $C$.  

Now, we will model Alice and Bob's choice of $\MR$ in the prepare and measure protocol as them applying a CNOT operation to a private register of size $N$ qubits each and then later measuring their private register in the $Z$ basis.  If they choose $\R$, they will not apply this CNOT register.  It is not difficult to see that this is equivalent to the real protocol where they would measure immediately.  Let $\Theta_A$ and $\Theta_B$ be their choice of $\R$ or $\MR$ where $\Theta_A,\Theta_B \in \{0,1\}^N$ and a $1$ in index $i$ indicates a choice to $\MR$ signal $i$ (i.e., apply a CNOT gate).

In more detail, Alice and Bob will start the protocol with a register of size $N$ qubits each, cleared to the all zero state $\ket{0\cdots 0}$.  On round $i$, if $\Theta_A^i=0$, Alice will apply the identity operator to qubit $i$; if $\Theta_A^i = 1$, Alice will apply a CNOT gate with the control register being the $i$'th qubit sent from the server, and the target register being the $i$'th ancilla qubit in Alice's private register.  Similarly for Bob.  Thus, we can actually write the result of this operation on a bit string $\ket{a} = \ket{a_1\cdots a_N}$ to be:
\begin{equation}
  \ket{0\cdots 0}_A\otimes \ket{a}_{T_1} \mapsto \ket{a\wedge\Theta_A}\otimes\ket{a}_{T_1},
\end{equation}
where $a \wedge \Theta_A$ is the bit-wise logical AND, namely $a \wedge \Theta_A = (a_1 \wedge \Theta_A^1)\cdots (a_N\wedge \Theta_A^N)$.  The fact that we can write this as a logical AND is due to the fact that the ancilla is cleared to zero; thus it changes to a $\ket{1}$ only if the corresponding bit in $a$ and $\Theta_A$ are both one.

A general attack against the prepare-and-measure protocol will consist of the adversarial server preparing an arbitrary $2N$-qubit initial state of the form:
\begin{equation}
    \ket{\psi_0} = \sum_{i,j\in\{0,1\}^N}\alpha_{i,j}\ket{i,j,c_{i,j}}_{T_1T_2E}.
\end{equation}
Alice and Bob will receive the $T_1$ and $T_2$ qubits respectively (while the server keeps the $E$ system private).  The users will then apply a CNOT gate to their respective $T$ register and a system held private by each user as discussed.  After this, the state becomes:
\begin{equation}
    \ket{\psi_1} = \sum_{i,j\in\{0,1\}^N}\alpha_{i,j}\ket{i\wedge\Theta_A,j\wedge\Theta_B}\otimes\ket{i,j,c_{i,j}}_{T_1T_2E},
\end{equation}
The $T$ registers return to the adversarial server who applies a quantum instrument which, through standard techniques, may be dilated to a unitary operator as before (though now, of course, this operator acts on all $2N$ qubits and the $E$ register).  This operator $U$ will act, without loss of generality, as follows:
\[
U\ket{i,j,c_{i,j}} = \sum_{m\in\{0,1,2,3\}^N}\ket{m}_C\ket{f_{i,j}^m}.
\]
Note that $U$'s action on states of the form $\ket{i,j,c_{a,b}}$ for $(i,j) \ne (a,b)$ can be arbitrary as they do not appear in the returned state.  The resulting state, then, is:
\begin{equation}\label{eq:final-p-and-m}
    \ket{\psi_2} = \sum_{i,j\in\{0,1\}^N}\alpha_{i,j}\ket{i\wedge\Theta_A,j\wedge\Theta_B}\otimes\sum_{m\in\{0,\cdots,3\}^N}\ket{m}_C\ket{f_{i,j}^m}_E.
\end{equation}
Following this, the server would measure the message register which dictates the server's message and post-measurement state.

At this point, let us consider the entanglement based version and show there is an attack that the adversarial source may use which produces the exact same state as Equation \ref{eq:final-p-and-m}, conditioning on Alice and Bob not aborting.  First, Eve will prepare the state:
\begin{equation}\label{eq:ent-state}
    \ket{\zeta_0} = \sum_{i,j\in\{0,1\}^N}\alpha_{i,j}\ket{i,j}_{AB}\otimes\sum_{m\in\{0,\cdots, 3\}^N}\ket{m}_C\ket{f_{i,j}^m}_C.
\end{equation}
Clearly this is something that Eve can prepare.  Indeed, she could initially prepare the state $\sum_{i,j}\alpha_{i,j}\ket{i,j}_{AB}\ket{i,j}_{T_1T_2}\ket{c_{i,j}}_E$ and then apply $U$ to the right-most two registers which will evolve the state to the above.  She sends the $A$ and $B$ registers to Alice and Bob respectively while sending the $C$ register to the trusted server. We claim this is the desired state; namely that if Alice and Bob both observe a ``+'' on the systems where $\Theta_A$ and $\Theta_B$ are $0$, the collapsed state is identical to Equation \ref{eq:final-p-and-m}.

For a given $\Theta_A$ and bit string $i$, we may decompose $i$ into a ``zero'' part (those indices of $i$ where $\Theta_A$ is a zero) and a ``one'' part (those indices of $i$ where $\Theta_A$ is a one).   Then, there exists a natural permutation $\pi_A$ such that every string $i$ can be written as $i = \pi_A(i_0, i_1)$ and $i\wedge\Theta_A = \pi_A(0\cdots 0, i_1)$.  Similarly for $B$ (with permutation $\pi_B$).  For example, if $\Theta_A = 0110010$ then $\pi_A(x_1x_2x_3x_4,y_1y_2y_3) = x_1y_1y_2x_2x_3y_3x_4$ and, furthermore, if given $i=1001011$, then $i_0 = 1101$ (those parts of $i$ that match with a zero in $\Theta_A$) and $i_1 = 001$.  In this case $\pi_A(i_0,i_1) = \pi_A(1101,001) = 1001011 = i$.

Finally, let $c_0(x)$ be the number of $0$'s in the bit-string $x$; similarly define $c_1(x)$ to be the number of $1$'s in the bit string $x$.  From this, we may write Equation \ref{eq:final-p-and-m} as follows:
\begin{align}
  \ket{\psi_2} &= \sum_{\substack{i_0\in\{0,1\}^{c_0(\Theta_A)}\\i_1\in\{0,1\}^{c_1(\Theta_A)}\\j_0\in\{0,1\}^{c_0(\Theta_B)}\\j_1\in\{0,1\}^{c_1(\Theta_B)}}}\alpha_{i,j}\ket{\pi_A(0,i_1)}_A\ket{\pi_B(0,j_1)}_B\otimes\sum_{m\in\{0,\cdots,3\}^N}\ket{m}_C\ket{f_{i,j}^m}_E\notag\\\notag\\
  &= \sum_{\substack{i_1\in\{0,1\}^{c_1(\Theta_A)}\\j_1\in\{0,1\}^{c_1(\Theta_B)}}}\ket{\pi_A(0,i_1)}_A\ket{\pi_B(0,j_1)}_B\otimes\underbrace{\sum_{\substack{i_0\in\{0,1\}^{c_0(\Theta_A)}\\j_0\in\{0,1\}^{c_0(\Theta_B)}}}\alpha_{i,j}\sum_{m\in\{0,\cdots,3\}^N}\ket{m}_C\ket{f_{i,j}^m}_E}_{\ket{\phi(i_1,j_1)}_{CE}}\notag\\\notag\\
    &= \sum_{\substack{i_1\in\{0,1\}^{c_1(\Theta_A)}\\j_1\in\{0,1\}^{c_1(\Theta_B)}}}\ket{\pi_A(0,i_1)}_A\ket{\pi_B(0,j_1)}_B\otimes\ket{\phi(i_1,j_1)}_{CE}.\label{eq:p-and-m-final2}
\end{align}
Note that, since the above state is normalized (due to the unitarity of the attack operations), it holds that:
\begin{equation}\label{eq:normalized}
  \sum_{\substack{i_1\in\{0,1\}^{c_1(\Theta_A)}\\j_1\in\{0,1\}^{c_1(\Theta_B)}}}\bk{\phi(i_1,j_1)} = 1.
\end{equation}

Now, consider the state created by Eve for the entanglement based protocol, namely Equation \ref{eq:ent-state}.  Using the same function $\pi_A$ and $\pi_B$, we may write it as:
\begin{align}
  \ket{\zeta_0} &= \sum_{\substack{i_0\in\{0,1\}^{c_0(\Theta_A)}\\i_1\in\{0,1\}^{c_1(\Theta_A)}\\j_0\in\{0,1\}^{c_0(\Theta_A)}\\j_1\in\{0,1\}^{c_1(\Theta_B)}}}\alpha_{i,j}\ket{\pi_A(i_0,i_1)}_A\ket{\pi_B(i_1,j_1)}_B\otimes\sum_{m\in\{0,\cdots,3\}^N}\ket{m}_C\ket{f_{i,j}^m}_E.
\end{align}
We now change basis for those qubits of $A$ where $\Theta_A$ is a zero (also for those qubits in $B$'s register).  We write $\pi_A(+,i_1)$ to be the function which places a character ``$+$'' in the output string wherever $\Theta_A = 0$ (same for $\pi_B$).  This allows us to write $\ket{\zeta_0}$ as:
\begin{align}
  \ket{\zeta_0}&= \frac{1}{M}\sum_{\substack{i_1\in\{0,1\}^{c_1(\Theta_A)}\\j_1\in\{0,1\}^{c_1(\Theta_B)}}}\ket{\pi_A(+,i_1)}_A\ket{\pi_B(+,j_1)}_B\otimes\sum_{\substack{i_0\in\{0,1\}^{c_0(\Theta_A)}\\j_0\in\{0,1\}^{c_0(\Theta_B)}}}\alpha_{i,j}\sum_{m\in\{0,\cdots,3\}^N}\ket{m}_C\ket{f_{i,j}^m}_E + \ket{\nu}_{ABCE}\notag\\\notag\\
  &=\frac{1}{M}\sum_{\substack{i_1\in\{0,1\}^{c_1(\Theta_A)}\\j_1\in\{0,1\}^{c_1(\Theta_B)}}}\ket{\pi_A(+,i_1)}_A\ket{\pi_B(+,j_1)}_B\otimes\ket{\phi(i_1,j_1)} + \ket{\nu}_{ABCE}\notag\\\notag\\
\end{align}
where $M = \sqrt{2^{c_0(\Theta_A)}}\sqrt{2^{c_0(\Theta_B)}} > 0$ and $\ket{\nu}_{ABCE}$ is some quantum state where the $A$ or $B$ registers contain at least one $\ket{-}$ in a position where the corresponding $\Theta_{A/B}$ is a zero.  That is $\ket{\nu}_{ABCE}$ is a state which would cause an abort of the protocol.  From the above, it is clear that, conditioning on \emph{not} aborting, the state collapses to Equation \ref{eq:p-and-m-final2}, the actual state resulting from the prepare-and-measure semi-quantum protocol.  Furthermore, it is clear from Equation \ref{eq:normalized}, along with $M>0$, that the probability of not aborting is strictly positive.  This completes the reduction.  Thus, since, following this stage, the two protocols are identical, the claim follows and so proving security against general attacks in the entanglement based protocol will imply security against general attacks for the real prepare-and-measure version; that is, $\texttt{Gen-Ent} \Longrightarrow \texttt{Gen-PM}$.

Clearly this entanglement based protocol is not an efficient one, however that does not matter.  Instead, we are showing that for any attack against the prepare and measure protocol (which never aborts unless the noise is too high) there exists an attack against the entanglement one such that (1) the quantum states, conditioning on a non-abort of the entanglement protocol, are identical in both protocols (thus any key-rate computation in one will apply to the other); (2) this equivalent attack always has a non-zero probability of not aborting (so that we are not analyzing any cases in the entanglement protocol that cannot occur in order to prove security of the prepare and measure protocol).  Note that there are attacks against the entanglement protocol, as discussed, which always abort; however from the above, those attacks do not show up in the prepare and measure version and so are not worth analyzing (as we care only about the prepare and measure protocol).  Note also that there may be more attacks against the entanglement protocol; thus if the entanglement protocol is secure with a positive key rate, the prepare and measure protocol will also be secure with a key rate at least as high (possibly higher).

Using de Finetti or postselection style techniques \cite{konig2005finetti,christandl2009postselection} it can be shown that security against collective attacks in the entanglement based version imply security against general attacks in the entanglement based version (i.e., $\texttt{Col-Ent} \Longrightarrow \texttt{Gen-Ent}$).  Indeed, the entanglement based version may be made permutation invariant in the standard way by having $A$ and $B$ permute their subsystems \cite{QKD-renner-keyrate}.  Thus, we may assume that the state Eve prepares initially in the entanglement based protocol is of the form $\ket{\zeta_0} = \ket{\mu_0}^{\otimes N}$, where we may write $\ket{\mu_0}$ in the most general way as:
\[
\ket{\mu_0} = \sum_{i,j\in\{0,1\}}\alpha_{i,j}\ket{i,j}_{AB}\otimes\sum_{m\in\{0,\cdots,3\}}\ket{m}_C\otimes\ket{e_{i,j}^m}.
\]

It is easy to see that, regardless of Alice and Bob's choice of $\MR$ or $\R$ at this point (conditioning on a non-abort),  the state will be equivalent to the one we analyzed in the previous section.  Namely, for any initial state of the above form for the entanglement based protocol (that is, $\ket{\mu_0}$), there exists an initial state and return attack operator $U$ in the prepare and measure case which will match the above expression.  Thus, our entropy bound will apply in this case and so $\texttt{Col-PM} \Longrightarrow \texttt{Col-Ent}$.  Therefore, taken together, we may conclude that our prepare-and-measure protocol is actually secure against general attacks.  Note that the above analysis may be useful for the proofs of security of other (S)QKD protocols, mediated or otherwise.

\subsubsection{Interactive Attacks}
We now show the reduction to an entanglement based protocol for the more complex scenario where the adversary is allowed adaptive, interactive, attacks.  By this, we mean the adversary can send a qubit to Alice or Bob, and wait for the qubit back before deciding what to send next.  Furthermore, the systems may be out of order (e.g., the adversary may first send qubits to Alice and then adjust its attack before sending qubits to Bob).  This reduction technique we develop here may be useful in other two party protocols relying on two-way quantum channels outside this single, particular, semi-quantum protocol we are analyzing in this work.

\begin{figure}
    \centering
    \includegraphics[width=.5\textwidth]{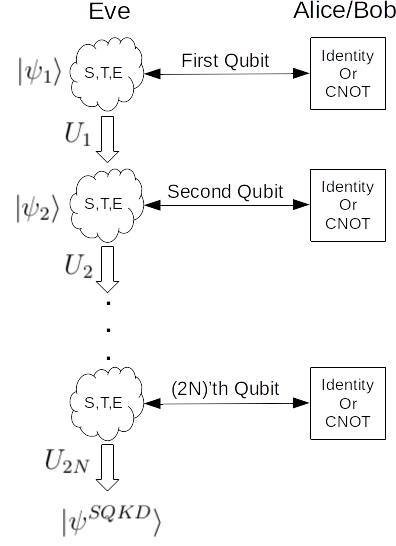}
    \caption{Showing the interactive attack model we consider in this section.  Eve is allowed to prepare an arbitrary initial state consisting of $2N$ qubits (the Transit, $T$, register); an ordering decision of who to send qubits in which order (the Selection, $S$, register); and an entangled ancilla (the $E$ register).  Eve then sends the first qubit to a party of her choice.  That party performs their given operation and returns the qubit to Eve.  Eve is then allowed to perform an arbitrary unitary operation on all $2N$ qubits, the selection choice, and her private ancilla (i.e., she may adapt her attack in the second round based on the response from the first).  A second qubit is then sent to a party of Eve's choice.  This repeats until all $2N$ qubits have been sent and received leading to the final quantum state denoted $\ket{\psi^{SQKD}}$.  We claim in this section that an equivalent state may be created for the entanglement based protocol thus implying that security of the entanglement based protocol will imply security of the semi-quantum one.}
    \label{fig:interactive-attack}
\end{figure}

In more detail, for this attack, Eve is allowed to first create an arbitrary $2N$-qubit state, possibly entangled with Eve's private ancilla, where $N$, as before, is the number of rounds the protocol will use.  Eve will then decide, potentially through some probabilistic process, which party, Alice or Bob, to send the first qubit to.  That party, on receipt of the qubit will perform their choice of operation ($\MR$ or $\R$) and return the qubit to the adversary.  Eve is now allowed to probe the entire $2N$-qubit state, along with her entangled ancilla, to evolve the system to a new $2N$-qubit state, again entangled with Eve's private memory.  Eve then chooses a party to send the second qubit to.  This process repeats until  $2N$ qubits have been transmitted and received back.  Finally, Eve applies a quantum instrument (which, as before, will be dilated to a unitary operator) which determines the classical message she sends (this message being in the set $\{0, 1, 2, 3\}^{N}$) and her post measured ancilla state.  See Figure \ref{fig:interactive-attack}.

Note that, despite the allowed adaptive interactivity, we still assume ideal qubits and, furthermore, we assume Eve sends exactly $N$ qubits to Alice and $N$ qubits to Bob.  Of course, there are several more practical attacks which are outside our security model - for instance, Eve could send two photons to a party when that party expects only one; if the party chooses $\MR$, thus destroying the photon, information may be leaked to Eve.  Such attacks are outside the scope of our security model (though we comment on them in the next sub-section).  We assume, here, ideal qubits and that when Alice or Bob receives quantum bits from Eve, they can operate on them individually.  Nonetheless, Eve has a lot of flexibility in creating the quantum state and adapting her attack based on users' actions throughout the protocol.

We first consider the prepare-and-measure semi-quantum protocol and model the attack.  We will then show, as in the previous sub-section for non-interactive attacks, that an equivalent state may be prepared for the entanglement based protocol, leading to the same quantum state after Alice and Bob's operations, conditioning on a non-abort.  We begin by having a Transit register ($T$) of $2N$ qubits, which will represent the qubits being sent to Alice and Bob, and Eve's ancilla, similar to before.  We also have a private ancilla for Alice and Bob of size $2N$ qubits which, as before, will be used to model their classical memory for measurements.

We now introduce a new register of size $2N$-qubits which will be used to decide which party to send the next qubit to.  In particular, if the $i$'th qubit of this register is a $\ket{0}$, the $i$'th qubit should go to Alice; otherwise it should go to Bob.  We use a quantum register here as it will allow us to model probabilistic strategies and also adaptive ones (where the register can be altered after each qubit received back).  We will call this the Selection ($S$) register.  The initial state Eve prepares, then, may be written as:
\begin{equation}
  \ket{\psi_1} = \sum_{s^1\in\{0,1\}^{2N}}\sum_{t^1\in\{0,1\}^{2N}}\ket{s^1,t^1}_{ST}\ket{E_{s^1t^1}}\otimes\ket{0\cdots0}_{AB}.
\end{equation}
Note that the $\ket{E_{s^1t^1}}$ states are not necessarily normalized.  For notation, we will use $t^i$ to mean the state of the transit register on round $i$ (starting at round $1$).  We will decompose $t^i$ as $t^i = t^i_1t^i_2\cdots t^i_{2N}$.  Similarly for the selection register $s^i$.

Note that the Selection register is in a superposition.  This models fixed order choice attacks (where Eve determines before the protocol runs, what order to send the qubits in); it also models probabilistic choices (as Eve can measure the register to determine what order to use).  We keep it as a superposition to give Eve the most flexibility.  In this event, Alice and Bob's operations on round $i$ are actually conditioned on the $i$'th qubit of the Selection register.

After creating the above state, the first qubit in the Transit register is sent out.  As before, we can model a $\MR$ operation as a CNOT gate, with the target being that users' private ancilla (initially cleared to $\ket{0}$) and the control being the transit register.  Note that we add a second control, namely the Selection register; if Alice wants to $\MR$, she will do so only if the corresponding Selection register is a $\ket{0}$; Bob will only do so if the Selection register is a $\ket{1}$.  Of course, in practice, Alice and Bob cannot access this register; however we are assuming ideal qubits and that users know when they receive a qubit.

We will assume that the first $N$ qubits of the $AB$ register belong to Alice and the second $N$ qubits belong to Bob.  Let $\pi^{s_1^1}(t^1_1, 0)$ be the function which ``places'' the bit $t_1^1$ in Alice's first private register if $s_1^1=0$; otherwise it places $t_1^1$ in Bob's first register if $s_1^1 = 1$.  Namely:
\[
\pi^{s_1^1}(t^1_1,0) = \left\{\begin{array}{ll}
t_1^1||0^{2N-1} & \text{ if } s_1^1 = 0\\
0^N||t_1^1||0^{N-1} & \text{ if } s_1^1 = 1\end{array}\right.
\]
where $0^x$ means a bit-string of size $x$ consisting of all zeros and the ``$||$'' operation is bit-string concatenation.  Finally, let $\Theta = \Theta_A||\Theta_B$.  Then, using the same arguments as in the previous sub-section when discussing the $\MR$ operation modeled as a CNOT gate, the state, after Alice and Bob's operation, evolves to:
\begin{equation}
  \ket{\psi_1'} = \sum_{s^1\in\{0,1\}^{2N}}\sum_{t^1\in\{0,1\}^{2N}}\ket{s^1,t^1}_{ST}\ket{E_{s^1t^1}}\otimes\ket{\pi^{s_1^1}(t_1^1,0)\wedge\Theta}_{AB}.
\end{equation}

The qubit returns to Eve who now has full control of the $S$, $T$, and $E$ registers.  She then applies a unitary probe $U_1$ to these registers, evolving them all (thus allowing her to change her ordering decision in the Selection register based on the response from the users).  Without loss of generality, we may define $U_1$'s action as follows:
\begin{equation}
  U_1\ket{s^1,t^1}\ket{E_{s^1t^1}} = \sum_{s^2\in\{0,1\}^{2N}}\sum_{t^2\in\{0,1\}^{2N}}\ket{s^2,t^2}_{ST}\otimes\ket{E_{s^2t^2|s^1t^1}}.
\end{equation}
Note that we need not define $U_1$'s action on states other than those above, as any other state will never appear in the quantum system under investigation.  Naturally, the states in Eve's ancilla are not normalized and unitarity of $U_1$ imposes constraints on them.  The state, after applying this probe, becomes:
\begin{equation}
  \ket{\psi_2} = \sum_{\substack{s^1\in\{0,1\}^{2N} \\ t^1\in\{0,1\}^{2N}}} \sum_{\substack{s^2\in\{0,1\}^{2N} \\ t^2\in\{0,1\}^{2N}}}\ket{s^2,t^2}\otimes\ket{E_{s^2t^2|s^1t^1}}\otimes\ket{\pi^{s_1^1}(t_1^1,0)\wedge\Theta}_{AB}
\end{equation}

The process then repeats with the second qubit of the $T$ register being sent to a party as determined by the second qubit of the $S$ register.  When that party returns the qubit to Eve, the state is in the form:
\begin{equation}
  \ket{\psi_2'} = \sum_{\substack{s^1\in\{0,1\}^{2N} \\ t^1\in\{0,1\}^{2N}}} \sum_{\substack{s^2\in\{0,1\}^{2N} \\ t^2\in\{0,1\}^{2N}}}\ket{s^2,t^2}\otimes\ket{E_{s^2t^2|s^1t^1}}\otimes\ket{\pi^{s_1^1s_2^2}(t_1^1t_2^2,0)\wedge\Theta}_{AB},
\end{equation}
where, above, we define the function $\pi^{s_1^1s_2^2}(t_1^1t_2^2,0)$ similarly to the one involving only the first round information; namely, it is a function that will place the bits $t_1^1$ and $t_2^2$ in the correct register of Alice and Bob based on the ordering $s_1^1s_2^2$.  For instance, if $s_1^1=s_2^2 = 0$, then the output will be $t_1^1t_2^2||0^{2N-2}$; or if $s_1^1 = 1$ and $s_2^2 = 0$, then the output will be $t_2^2||0^{N-1}||t_1^1||0^{N-1}$.  Eve, who again holds control of the complete $T$, $S$, and $E$ registers, will apply a new unitary probe $U_2$ whose actions we may define as:
\begin{equation}
  U_2\ket{s^2,t^2}_{ST}\otimes\ket{E_{s^2t^2|s^1t^1}} = \sum_{\substack{s^3\in\{0,1\}^{2N}\\t^3\in\{0,1\}^{2N}}}\ket{s^3,t^3}\otimes\ket{E_{s^3t^3|s^2t^2,s^1t^1}}.
\end{equation}
Similar to before, $U_2$'s action on states not of the form $\ket{s^2,t^2}_{ST}\otimes\ket{E_{s^2t^2|s^1t^1}}$ may be arbitrary as they do not appear.  Of course unitarity of $U_2$ places constraints on the (sub-normalized) $E$ states which will be important later.

This process repeats for $2N$ rounds.  After round $2N$, when the last party to act sends the $2N$'th qubit from the $T$ register back to Eve, but before Eve applies her quantum instrument, the state may be written in the form:
\begin{equation}
  \ket{\psi_{2N}'} = \sum_{\substack{s^1\in\{0,1\}^{2N} \\ t^1\in\{0,1\}^{2N}}}\cdots \sum_{\substack{s^{2N}\in\{0,1\}^{2N} \\ t^{2N}\in\{0,1\}^{2N}}}\ket{s^{2N},t^{2N}}\otimes\ket{E_{s^{2N}t^{2N}|s^{2N-1}t^{2N-1}\cdots s^1t^1}}\otimes\ket{\pi^{s_1^1\cdots s_{2N}^{2N}}(t_1^1\cdots t_{2N}^{2N})\wedge\Theta}_{AB},
\end{equation}

Now, Eve again controls the $S$, $T$, and $E$ registers and applies a quantum instrument.  As in the previous section, this can be dilated to an isometry $U_{2N}$ acting as follows:
\begin{equation}
  U_{2N}\ket{s^{2N},t^{2N}}_{ST}\otimes\ket{E_{s^{2N}t^{2N}|s^{2N-1}t^{2N-1}\cdots s^1t^1}} = \sum_{m\in\{0,\cdots,3\}^{N}}\ket{m}_C\otimes\ket{E_{m|s^{2N}t^{2N}\cdots s^1t^1}}_E.
\end{equation}
Note that, above, the $S$ and $T$ registers are absorbed into the final $E$ ancilla state.  Thus, the final state after running the actual semi-quantum protocol under this attack is found to be:
\begin{equation}
  \ket{\psi^{SQKD}} = \sum_{\substack{s^1\in\{0,1\}^{2N} \\ t^1\in\{0,1\}^{2N}}}\cdots \sum_{\substack{s^{2N}\in\{0,1\}^{2N} \\ t^{2N}\in\{0,1\}^{2N}}}\sum_{m\in\{0,\cdots,3\}^{2N}}\ket{m}\otimes\ket{E_{m|s^{2N}t^{2N}\cdots s^1t^1}}\otimes\ket{\pi^{s_1^1\cdots s_{2N}^{2N}}(t_1^1\cdots t_{2N}^{2N})\wedge\Theta}_{AB}
\end{equation}

We now manipulate the above state:
\begin{align*}
  &\ket{\psi^{SQKD}} = \sum_{\substack{s^1\in\{0,1\}^{2N} \\ t^1\in\{0,1\}^{2N}}}\cdots \sum_{\substack{s^{2N}\in\{0,1\}^{2N} \\ t^{2N}\in\{0,1\}^{2N}}}\sum_{m\in\{0,\cdots,3\}^{2N}}\ket{m}\otimes\ket{E_{m|s^{2N}t^{2N}\cdots s^1t^1}}\otimes\ket{\pi^{s_1^1\cdots s_{2N}^{2N}}(t_1^1\cdots t_{2N}^{2N})\wedge\Theta}_{AB}\\
  &\cong\sum_{\substack{s_1^1\in\{0,1\}\\s_2^2\in\{0,1\}\\\vdots\\s_{2N}^{2N}\in\{0,1\}}}\sum_{\substack{t_1^1\in\{0,1\}\\t_2^2\in\{0,1\}\\\vdots\\t_{2N}^{2N}\in\{0,1\}}}\ket{\pi^{s_1^1\cdots s_{2N}^{2N}}(t_1^1\cdots t_{2N}^{2N})}_{AB}\otimes\underbrace{\left(\sum_{\substack{s_{-1}^{1}\in\{0,1\}^{2N-1}\\\vdots\\s_{-2N}^{2N}\in\{0,1\}^{2N-1}}}\sum_{\substack{t_{-1}^{1}\in\{0,1\}^{2N}\\\vdots\\t_{-2N}^{2N}\in\{0,1\}^{2N-1}}}\sum_{m}\ket{m}\otimes\ket{E_{m|s^{2N}t^{2N}\cdots s^{1}t^{1}}}\right)}_{\ket{\phi(s_1^1\cdots s_{2N}^{2N}, t_1^1\cdots t_{2N}^{2N})}}
\end{align*}
Observe that the left-most summations are over single bits $s_i^i$ and $t_i^i$ whereas the summations on the right are over the remaining $2N-1$ bits of those respective strings.  We use the notation $s_{-i}^{i}$ to mean the substring of $s^i$ that does not include the $i$'th bit (i.e., $s^1 = s^1_1||s^1_{-1}$).  We also permuted the subspaces at this point, putting the $AB$ register on the left, for clarity only.  Changing notation slightly, we may write the above more simply as:
\begin{align}
  \ket{\psi^{SQKD}} &\cong \sum_{s\in\{0,1\}^{2N}}\sum_{t\in\{0,1\}^{2N}}\ket{\pi^{s_1\cdots s_{2N}}(t_1\cdots t_{2N})\wedge \Theta}_{AB} \otimes \ket{\phi(s,t)}\notag\\
  &= \sum_{t\in\{0,1\}^{2N}}\ket{t\wedge\Theta}_{AB}\otimes\sum_{\substack{s\in\{0,1\}^{2N}\\u  \st \pi^s(u) = t}}\ket{\phi(s,u)}.\label{eq:final-int-sqkd}
\end{align}

Let us now consider the entanglement-based protocol.  Here, Eve is allowed no interactivity and must create a single quantum state, sending part to Alice, part to Bob, and part to a trusted server while keeping the remainder for herself.  We show that there is an initial state that Eve can create which exactly mimics the above state, assuming Alice and Bob do not abort the entanglement based protocol.  Furthermore, we show that this created state has a non-zero probability of not aborting.  We claim the desired state can be created by Eve by simulating the semi-quantum protocol, playing the part of Alice and Bob, but simulating the case when both Alice and Bob always choose $\MR$ (i.e., when $\Theta = 1\cdots 1 = 1^{2N}$).  Such a state can clearly be created by Eve and the resulting state is found to be (using the simplified notation above):
\begin{align}
  \ket{\psi^{ent}} &\cong \sum_{s\in\{0,1\}^{2N}}\sum_{t\in\{0,1\}^{2N}}\ket{\pi^{s_1\cdots s_{2N}}(t_1\cdots t_{2N})}_{AB} \otimes \ket{\phi(s,t)}\notag\\
  &=\sum_{t\in\{0,1\}^{2N}}\ket{t}_{AB}\otimes\sum_{\substack{s\in\{0,1\}^{2N}\\u  \st \pi^s(u) = t}}\ket{\phi(s,u)}.
\end{align}
At this point, we may use the same technique as in the non-interactive case to show that, conditioning on a non-abort of the entanglement based protocol (namely that Alice and Bob observe all $\ket{+}$ in their given registers when $\Theta = 0$), the post measurement state collapses exactly to the state produced by the actual semi-quantum protocol (Equation \ref{eq:final-int-sqkd}).  Also using the same analysis above, taking into account that the attack operators are unitary, the probability of not aborting is strictly positive.  This completes the analysis.

\subsection{Comment on Practical Attacks and Implementations}

Our work in this section has focused on the theoretical, ideal device and single qubit, scenario.  We showed that an improvement in efficiency is possible under these conditions based on the noise level of the channel, however it is worth discussing practical considerations.  When implementing a QKD protocol, one often uses weak coherent sources \cite{QKD-survey1} which produce, with non-zero probability, vacuum states or, often worse from a security stand point, multi-photon states.  These multi-photon states open up attacks such as photon number splitting attacks \cite{photon-split1,photon-split2}.  Such attacks are often mitigated using decoy-state methods \cite{decoy-0,decoy-1,decoy-2,decoy-rate}; though it is an open question whether or not those methods can help in the semi-quantum scenario.  In the semi-quantum case, however, things are even more challenging.  Due to the two-way channel and the use of the $\MR$ operation, Eve is afforded even more attack opportunities, such as the photon-tagging attack \cite{SQKD-photon-tag,SQKD-photon-tag-comment}.  In general, any semi-quantum protocol implementing the $\MR$ operation cannot be experimentally feasible \cite{boyer2017experimentally}; however, one can modify the $\MR$ operation using ``mirror-devices'' as proposed in \cite{boyer2017experimentally}, however this comes at the cost of reducing efficiency by at least $50\%$.  Indeed, as shown in \cite{massa2019experimental}, a M-SQKD protocol was proven secure assuming practical devices, but with a key-rate of only $12.5\%$ in the ideal scenario and less than $1\%$ using practical current-day devices.  Though we leave this as an open problem, we suspect our protocol can be implemented using mirror-style devices as in \cite{boyer2017experimentally}, though with a similar drop in efficiency.

Despite these short-comings when translating theoretical semi-quantum results to practice, we still feel the study of semi-quantum cryptography is of high importance.  First, due to the increased complexity of the attacks against them (due to the two-way channel and also due to users' device restrictions), standard security proof techniques often fail and so new methods are required.  These new methods can lead to new insights and new mathematical tools for other researchers to apply to alternative QKD protocols which may actually be more practical.  The design of SQKD protocols also requires careful use of channel statistics, such as the use of mismatched measurements - these insights can be valuable in other QKD research.  Finally, it also addresses fundamental questions providing us with insight into the ``gap'' between classical and quantum communication.

\section{Closing Remarks}

In this paper, we extended the original M-SQKD paper from \cite{krawec2015mediated} to improve efficiency.  Our modifications allow for nearly doubling of the key-generation rates for low noise levels.  Though this comes at the cost of reduced noise tolerance, our protocol was designed to be ``backwards compatible'' with the original M-SQKD protocol.  In fact, users may even decide \emph{after} the quantum communication stage is finished whether to run the new, modified protocol or the original. Thus, taken together, our work shows how improved efficiency is possible for certain channel noise levels, without sacrificing noise tolerance as the users may switch to the original protocol if the observed noise level is too high.  While newer M-SQKD protocols, as discussed earlier, also exist now with asymptotically perfect efficiency \cite{m-2-liu2018mediated,m-3-lin2019mediated,m-9-chen2021efficient,m-10-hwang2020mediated}, ours is the first such protocol with provable security.

Towards the security proof, we also showed how this M-SQKD protocol, involving two-way quantum communication with an adversarial server, may be reduced to an entanglement based protocol.  This is perhaps the largest contribution of this work and our techniques here may be useful in other (S)QKD protocols involving two-way quantum communication.  Using this reduction, we were able to show a complete security analysis against general attacks.  Our methods here may be broadly applicable to other multi-user QKD protocols including and beyond M-SQKD ones.  In particular, our proof methods might be useful in proving the security of other M-SQKD protocols which have yet to obtain a key rate derivation.

Many interesting future problems remain open.  In particular, we did not consider practical attacks.  Due to the nature of the $\MR$ operation, several attack strategies against practical devices \cite{SQKD-photon-tag,SQKD-photon-tag-comment} are open which were not part of our security model (which assumed ideal qubit states).  Methods from \cite{boyer2017experimentally,massa2019experimental}, combined with our new security proof method (specifically our reduction to an entanglement based protocol), may create a more practical system.  Using our methods to prove the security of other M-SQKD protocols would also be very useful and allow us to compare the overall efficiency under noise of these many M-SQKD protocols in existence today to determine which M-SQKD protocol is actually the most efficient over a given quantum channel (e.g., attack scenario).

%\bibliographystyle{unsrt}
%\bibliography{local}

\end{document}